\def\kms{\ifmmode{\hbox{km~s}^{-1}}\else{km~s$^{-1}$}\fi}
\def\la{\mathrel{\hbox{\rlap{\hbox{\lower4pt\hbox{$\sim$}}}\raise1pt\hbox{$<$}}}
}
\def\ga{\mathrel{\hbox{\rlap{\hbox{\lower4pt\hbox{$\sim$}}}\raise1pt\hbox{$>$}}}
}
\def\k'{$K^{\prime}$}
\def\mk{$M_{K^{\prime}}$}
\def\deg{^{\circ}}
\documentstyle[12pt,aasms4,tighten]{article}
\begin{document}
\title {\bf Global Extinction in Spiral Galaxies}

\author{R. Brent Tully$^1$, Michael J. Pierce$^2$,
Jia-Sheng Huang$^1$, Will Saunders$^3$, 
Marc A.W. Verheijen$^4$, Peter L. Witchalls$^{3,5}$}

\affil {$^1$ Institute for Astronomy, University of Hawaii, Honolulu, HI
96822}
\affil {$^2$ Department of Astronomy, Indiana University}
\affil {$^3$ Institute for Astronomy, University of Edinburgh}
\affil {$^4$ Kapteyn Astronomical Institute, University of Groningen} 
\affil {$^5$ Deceased}

\begin{abstract}
Magnitude--limited samples of spiral galaxies drawn from the Ursa
Major and Pisces clusters are used to determine their extinction
properties as a function of inclination.  Imaging photometry is
available for 87 spirals in $B,R,I$, and \k' bands.  Extinction causes
systematic scatter in color--magnitude plots.  A strong luminosity
dependence is found.  Relative edge-on to face-on extinction of
up to $1.7^m$ is found at $B$ for the most luminous galaxies
but is unmeasurably
small for faint galaxies.
At $R$ the differential absorption with inclination reaches $1.3^m$,
at $I$ it reaches $1.0^m$, and at \k' the differential absorption
can in the extreme be as great as
$0.3^m$.  

The luminosity dependence of reddening can be translated into a 
dependence on rotation rate which is a distance-independent observable.
Hence, corrections can be made that are useful for distance
measurements.  The strong dependence of the corrections on luminosity 
act to steepen luminosity-linewidth correlations.  The effect is greatest 
toward the blue, with the consequence that luminosity--linewidth slope 
dependencies are now only weakly a function of color.

\end{abstract}

\keywords{galaxies: photometry - galaxies: ISM}

\section{Introduction}

It is evident that there is optical obscuration in spiral
galaxies yet the quantitative amount of the dimming due to obscuration
is disputed.  There is a displacement in the photometric properties of
edge-on and face-on galaxies.  If distance considerations are taken
into account, then in a set of similarly sized galaxies the more edge-on
ones will tend to be fainter, or in a set of galaxies of the same
luminosity the more edge-on ones will tend to be bigger.

This opening statement touches on the two outstanding problems that confront
efforts to quantify the level of obscuration in galaxies.  To begin with,
there is more than one good explanation for the photometric
separation of galaxies with inclination: galaxies of the same
intrinsic size and luminosity might appear dimmer toward edge-on
because of increased path lengths through the obscuring material, or
they might appear larger toward edge-on because surface brightnesses
are enhanced by the increased path lengths.  The second big problem
arises out of uncertain relative distances with most samples since 
luminosities and dimensions scale differently with distance, hence
distance uncertainties add noise to tests.

A nice review of previous research is provided by Huizinga 
(1994)\markcite{h94}.  At one extreme, galaxies were accepted to be 
completely transparent and corrections were made only to dimensions 
in the {\it Reference Catalogue of Bright Galaxies} (de Vaucouleurs
\& de Vaucouleurs 1964)\markcite{rc1}.  At the other extreme, galaxies
were proposed to be optically thick at all observable radii by Valentijn
(1990)\markcite{v90}, to the extent that Gonz\'alez-Serrano \& Valentijn
(1991)\markcite{gv91} posited the obscuring material could resolve the
`dark matter' problem.  The current generally accepted viewpoint is 
intermediate.  Spiral galaxies probably have high opacities near their 
centers but are relatively transparent at their visible extremities.

We have a utilitarian need to understand the global obscuration
properties of spiral galaxies.  A primary consideration for us is the
need to correct for inclination effects with luminosity -- HI profile
linewidth distance estimators (Tully \& Fisher 1977\markcite{tf}).
Inclination 
dependencies also have to be understood when we use photometric
information in statistical studies of galactic structure (cf, Tully \&
Verheijen 1997\markcite{tv}).  In these cases, it may be enough to
know the overall 
and statistical effects of projection from an empirical evaluation.
An alternative approach is to try to develop a physical model of what
is happening.  However, a realistic model would have a lot of parameters
and require a lot of information as constraints.  There have been
attempts in this direction (Wainscoat, Freeman, \& Hyland
1989\markcite{wfh}; Byun, Freeman, \& Kylafis 1994\markcite{bfk};
Bianchi, Ferrara, \& Giovanardi 1996\markcite{bfg}).  To 
date, there has 
not been any study with both the information content per object needed
for elaborate modeling and the large number of objects needed for a
statistical evaluation of the norm and range of obscuration
properties.

What we bring that is new is imaging photometry over a substantial
optical--infrared baseline: large format $B,R,I$ CCD imaging and
\k' imaging with a 256x256 HgCdTe detector.  Our sample is not really
large but it has nice qualities with respect to the
distance problem and with respect to completion.  We are using data 
from two clusters, hence objects
are at either of two discrete distances, and there is a measure of
completion to well specified limits.

Our primary sample is drawn from the Ursa Major Cluster.  The
photometric data has been published by Tully et al. (1996)\markcite{t5}.  
There are  
62 galaxies in a window on the sky and in redshift that are brighter 
than $14.7^m$
at $B$.  Optical photometry is available for all 62 and \k' photometry
is available for 60.  Observations and reductions of the optical
components were done by MJP, MAWV, and RBT.  Observations and
reductions of the \k' material were carried out by JSH, MAWV, and RBT.
The Ursa Major Cluster is a loose, irregular cluster with normal
spirals and few gas-poor systems.  It is probably dynamically young.
The constituents may be representative of typical galaxies outside of
dense regions.  Our sample is complete down to roughly the luminosity
of the Small Magellanic Cloud at $M_B \simeq -16.5^m$.  Unfortunately, 
there is a paucity of extremely luminous galaxies in this sample.

Our supplemental sample is drawn from what has been called the Pisces
Cluster, part of the Perseus-Pisces filament (Haynes \& Giovanelli
1988\markcite{hg88}), and an entity that has 
frequently been used in studies of distances and flows (Aaronson et
al. 1986\markcite{a86}; Han \& Mould 1992\markcite{hm}).
Sakai, Giovanelli, \& Wegner (1994)\markcite{sgw} have
discussed an overlapping part of 
the region and refer to subunits as the NGC~383 and NGC~507 groups.
We consider candidates within the window $00^h 49^m < \alpha < 01^h 32^m$,
$28\deg < \delta < 34\deg$, and 
$4,300~\kms < V_{helio}+300{\rm sin}\ell{\rm cos}b < 6,400~\kms$.  
There are a couple of knots in the region dominated by early-type
galaxies but spirals are scattered about in an irregular fashion.
We take galaxies detected in the HI 21 cm line at Arecibo (Giovanelli
\& Haynes 1985\markcite{gh85}, 1989\markcite{gh89}, Wegner, Haynes, \& 
Giovanelli 1993\markcite{whg}), typed
as spirals, and with cataloged axial ratios $b/a \le 0.5$.  The
Arecibo Telescope gives a high detection rate for spirals at 5,000~\kms\ 
so there is
reasonable completion (at $\delta < 33.5\deg$) to a magnitude limit
of $B = 15.7^m$.  This limit is $M_B = -19^m$ after corrections,
roughly the magnitude of the Local Group galaxy M33.
Cases where the HI signals may be
confused are avoided.  This Pisces sample has the disadvantages that
our axial ratio limit excludes
very face-on examples and for observational reasons it is not as 
rigorously complete as
the Ursa Major sample (details in next section).  However, it 
contains a substantial number of
intrinsically luminous spirals.  The optical photometry was acquired
and reduced by MJP and RBT.  The \k' photometry was obtained by PLW,
WS, and RBT.  There are 38 galaxies in the Pisces sample.

The information we have available allows us to excape the problems
that have plagued efforts to calibrate the levels of
obscuration in galaxies.  We will describe tests that decouple
luminosities and dimensions.  The cluster nature of the samples essentially
eliminates scatter due to uncertain relative distances.  The sample
completeness features are important.

Two different tests have been considered.  It turns out that the most
sensitive test involves deviations as a function of inclination from
the mean correlation in color-magnitude plots where one passband is
$B$, $R$, or $I$ and the other passband is \k'.   Another test gives
consistent but less statistically significant results.  It involves
deviations with inclination from the mean correlations between
luminosities and HI profile linewidths.  The latter test requires
kinematic information as well as photometric and cannot include very
face-on cases because of the large uncertainties in the deprojected
kinematic parameter.  Hence, fewer galaxies are available for the
second test and the inclination baseline is truncated.  Also, the
ratio of reddening scatter to the intrinsic luminosity scatter in 
the relations is
somewhat more favorable in the first test compared with the second.

Between the two clusters, we have 101 galaxies with $B,R,I,$\k'
information in our magnitude limited sample.  It will be seen that the
statistical effects of obscuration are quite evident.  The question
becomes, how complex a description of the obscuration does the data
afford?  The simplest case would be the definition of a single
coefficient for some analytic dependency on inclination or ellipticity
in each bandpass.  However, it has been suggested that there are
dependencies on such additional parameters as type (de Vaucouleurs et
al. 1991\markcite{rc3}; Han 1992\markcite{h92}) or
luminosity (Giovanelli et al. 1995).  Indeed, it will be shown that
there is a strong dependence of obscuration on luminosity in our
sample.  As Giovanelli et al. (1995) point out, the correction
procedure that is adopted affects the slope of the luminosity--linewidth 
correlation hence, potentially, distance estimates based on that
correlation.  The challenge to us is to generate a description of the
luminosity dependence with our relatively limited sample.  One
needs enough cases in a bin to sample a broad inclination range and
characterize the scatter.  Then the luminosity dependence has to be
transformed to a distance-independent parameter, like the profile
width, to be useful for distance measurements.

It would also be nice to further our understanding of how the
obscuring material is distributed in galaxies.  For example,
Giovanelli et al. (1994)\markcite{g94} and Peletier et al. 
(1995)\markcite{p95} contend that there is a strong radial
dependence, with considerable obscuration near the centers of galaxies
and little at the outer edges.  It would be possible to investigate
that kind of effect with two-dimensional color decompositions with our
data.  However, that study is beyond the scope of this paper.

\section{The Combined UMa and Pisces Data Sets}

The Ursa Major data set is the larger of the two cluster samples and
has more rigorous completion limits.  The cluster membership was
defined by windows on the plane of the sky and in velocity, as
discussed by Tully et al. (1996)\markcite{t5}.  The photometric data
are provided in that paper for the 79 galaxies accepted as cluster
members.  The census of the cluster within the proscribed window
constraints is considered to be complete to $B=14.7^m$, or after
preliminary corrections for inclination and assuming a distance
modulus of $31.33^m$ (18.5~Mpc), M$_B^{b,k,i} \sim -16.5^m$ at the 
faint limit.  Some
photometric, kinematic, and dynamical properties of these galaxies
were discussed by Tully \& Verheijen (1997)\markcite{tv} and 
Verheijen (1997)\markcite{v97}.

The distance to UMa was determined by the luminosity--linewidth
methodology (Pierce \& Tully, 1988\markcite{pt88},
1992\markcite{pt92}; Tully 1997).  For
intercomparisons within the UMa sample, the distance choice is
irrelevant; the only issue is the validity of the assumption that all
the distances are the same.  The intercomparisons {\it between} the
UMa and Pisces samples requires knowledge of the {\it relative}
distances of the two groups but the zero-point calibration of the
distance estimator is irrelevant.  Moreover, it will be shown that 
our primary color--magnitude test is quite {\it insensitive to modest 
relative distance
errors}.   Hence, the data are transformed to absolute
magnitudes to facilitate the intercomparison between clusters but 
the results are insensitive to uncertainties in distances.

All our color--magnitude comparisons are made with \k' magnitudes and
colors that stretch to \k' from one of $B,R$, or $I$.   This choice is
made because \k' is almost free of reddening hence, overwhelmingly,
the obscuration effects are found in the color term only.  The
$B$-band completion limit is found to translate to a \k' limit of
M$_{K^{\prime}}^{b,k}=-19.2^m$ for the UMa Cluster.  The superscripts 
indicate that corrections
have been made for reddening within our own Galaxy (latitude $b$) and for the
redshift spectral displacement ($k$-correction), although both are
negligible for 
the UMa systems.  These corrections are small but not completely
negligible for the Pisces systems to be discussed next.  The Galactic 
reddening corrections
are taken from Burstein \& Heiles (1984)\markcite{bh} in the following
ratios between bands: $B=1.00$, $R=0.62$, $I=0.41$, \k'$=0.08$.  The
spectral displacement adjustments ($k-$corrections) are $\sim 0$ for UMa 
galaxies and for Pisces galaxies are
$\sim 0.03^m\pm0.03$ at $B$ (Pence 1976)\markcite{p76} and $\la0.01^m$ at
$R$, $I$, and \k'.
The bands are
Johnson $B$, Cousins $R,I$, and a filtering of the $K$ infrared
atmospheric window that damps thermal noise.

After tiny Galactic-- and $k-$corrections (but {\it not} inclination
corrections) have been applied, there are 63 galaxies in the magnitude
limited sample (complete except for 2 galaxies missing \k' photometry:
the S$m$ galaxy UGC~6628 and the S$a$ galaxy UGC~7129).  Ten of these
are classified S0 or S0/$a$.  Our tests fail to demonstrate that there is
any extinction in this small sub-set of early-type galaxies.  Hence
these galaxies earlier than S$a$ are excluded from further
consideration.  The magnitude limit excludes any galaxy of type I$m$.
The UMa sample thus provides our analysis with 53 spirals typed 
S$a$ to S$m$.

Unfortunately, the UMa Cluster lacks extremely luminous members.  The
brightest galaxy at \k' is NGC~3953, an S$bc$ system with
$M_{K^{\prime}}^{b,k,i}=-24.35^m$, $M_B^{b,k,i}=-20.88^m$, and a maximum
rotation velocity $V_{max}=219$ \kms\ (these values have been 
corrected for inclination effects in ways to be described).  This
system is comparable to our Galaxy and considerably smaller than
Andromeda.  If extinction is a function of luminosity (Giovanelli et
al. 1995\markcite{g95}) then the sample needs to be extended to larger
systems.  The Pisces region sample suits our needs.

The Pisces Cluster galaxies selected for this study lie within the
space and velocity 
windows described in the previous section, have cataloged axial ratios
$b/a<0.5$ (nominally, inclinations greater than $60\deg$), are classed
as spirals, and have been detected 
in the HI radio line.   Giovanelli \& Haynes (1989\markcite{gh89})
claim `reasonable completion for spirals larger than $1^{\prime}$' as
reported in the {\it Uppsala General Catalogue} (Nelson
1973\markcite{n73}) and `more than 90\%' completion of spirals
brighter than $15.7^m$ in the {\it Catalogue of Galaxies and Clusters
of Galaxies} (Zwicky et al. 1961-68\markcite{z68}).  

The Pisces sample is taken to be at a distance of 60~Mpc,
$(m-M)_0=33.88^m$, $2.55^m$ more distant than UMa.  This distance is
determined from a recalibration of the luminosity--linewidth relations 
(Tully 1997) reviewed in Section~6.  The {\it differential} 
UMa--Pisces
distance comes from the superposition of the data from these two
regions.   It will be explained further along why our primary test is not
very sensitive to relative distance uncertainties.  At this assumed distance,
the equivalent \k'-band completion limit for our sample is
$M_{K^{\prime}}^{b,k}=-21.4^m$ (Galactic- and $k-$corrections applied but not
inclination corrections).  We have 4-band photometry for 34 galaxies
brighter than the completion limit.  The most luminous galaxy in this
Pisces sample has $M_{K^{\prime}}^{b,k,i}=-25.20^m$, $M_B^{b,k,i}=-21.79^m$,
and a maximum rotation velocity $V_{max}=259$~\kms, properties
resembling the Andromeda galaxy.

In summary, our analysis will be conducted with 87 galaxies; 53
spirals in UMa with $-24\la$\mk$^{b,k}\la-19$ and 34 spirals in Pisces
with $-25\la$\mk$^{b,k}\la-21$.  The UMa systems cover a full range of
inclinations but systems toward face-on were not observed in Pisces.

\section{Deviations from Color--Magnitude and Linewidth--Magnitude
Correlations} 

Color--magnitude diagrams are shown in Figure~1 for $B-$\k' vs. \mk,
$R-$\k' vs. \mk,  and $I-$\k' vs. \mk.  The different symbols
distinguish the UMa and Pisces samples.  Galactic- and $k-$corrections
have been applied but {\it not} corrections for obscuration as a function
of inclination.  The lines in Fig.~1 are regressions with
uncertainties in colors.  Obscuration as a function of tilt
essentially only affects the colors.

Our expectation is that more edge-on systems would be reddened by
extinction and tend to lie to the right in the color-magnitude
diagrams.  In fact, this expectation is met for the high luminosity
cases but is not clearly met for the faintest galaxies.  
Galaxies with \mk$^{b,k}<-23^m$ have considerable
extinction, and galaxies with \mk$^{b,k}>-20^m$ have negligible
extinction.  
The evidence is shown in Figure~2.  
After some experimentation, it
was decided the best we could do with the present information is to
separate the data into four luminosity zones.  
Deviations from the mean  $B-$\k',
$R-$\k', $I-$\k' vs. \mk\ fits are plotted as a function of axial
ratio in the four separate bins.

We adopt a standard empirical description of the
extinction, $A_{\lambda}$, as a function of inclination,
$i$, in the passband, $\lambda$:
\begin{equation}
A_{\lambda}^{i-0} = \gamma_{\lambda} {\rm log} (a/b).
\end{equation}
The nomenclature in the superscript, $i-0$, implies corrections are to
face-on orientation but do not account for extinction in a face-on
system.  Inclinations are related to axial ratios, $b/a$, through:
\begin{equation}
{\rm cos}i = \sqrt{{(b/a)^2-r_0^2 \over 1-r_0^2}}
\end{equation}
where $r_0$ is the axial ratio of a system viewed edge-on.  For the
purposes of this discussion we take $r_0=0.20$, although values as low
as $r_0 \sim 0.1$ can be entertained and the choice is not important
here.  Our corrections, 
$A_{\lambda}^{i-0}$, do not depend on the choice of $r_0$, only on
$b/a$.

Optimal values of $\gamma_{\lambda}$ were found for each of the four
luminosity bins and the three passbands $\lambda=B,R,I$.  The optimal
values satisfy the requirement that, with these corrections to
magnitudes, there are then no color--magnitude residuals with axial ratio.
The uncertainties that are recorded correspond to corrections that
give slopes which differ by one standard deviation from no correlation.
Account has to be taken that extinction is not entirely negligible at
\k'.  In this analysis, we assume $A_{K^{\prime}}$ is 15\% of $A_B$,
21\% of $A_R$, and 25\% of $A_I$.  These choices might sound surprisingly
high.  They were arrived at iteratively and will be justified in 
Section~5.

The best fits are shown in Fig.~2.  As expected, the amplitude of
extinction drops from $B$, through $R$, to $I$.  It is evident that
the amplitude of extinction within any of these bands also drops in
passing from high to low luminosity.  By the lowest luminosity bin
the correlation with axial ratio is so weak that it is not statistically
significant.  The information
in the three passbands is not entirely independent since the samples
are the same and all colors share the same \k'\ information.  These
results are summarized by the placement of the filled squares in
Figure~3. 

A similar analysis can be performed with the residuals of 
linewidth--magnitude correlations.  
This test is not quite as sensitive as the one already
described, for the reasons mentioned in the introduction.  Since fewer
galaxies are available, we consider three bins 
instead of four: $M_{K^{\prime}}<-23.4^m$ (21 galaxies),
$-23.4^m\leq M_{K^{\prime}}\leq -22.0^m$ (25 galaxies), and
$M_{K^{\prime}}>-22.0^m$ (16 galaxies).  The results from these tests
are recorded by the filled triangles in Fig.~3.

The evidence from the two tests are consistent and clearly indicate a
falling trend of $\gamma_{\lambda}$ toward the fainter galaxies.
Before we draw conclusions, though, let us compare these results with
what others have found.

\section{Comparison with Absorption Parameterizations in the
Literature}

Most of the early work involving large data sets use $B$ material, or
photographic magnitudes transformed to $B$.  
In the {\it Third Reference Catalogue} (de Vaucouleurs et
al. 1991)\markcite{rc3} extinction is described by the familiar
parameterization $A_B^{i-0}=\gamma_B{\rm log}(a/b)$, where
$\gamma_B$ is described as type dependent.  It
is a maximum value for type S$c$ at $\gamma_B=1.5$ and decreases
symmetrically with type steps toward earlier or later morphologies to
$\gamma_B=1.0$ at types 
S$a$ and S$m$.  There is a strong correlation between type and
luminosity, especially for S$bc$ and later (later types tend to be
fainter) so it can be expected that the alternative dependencies of
extinction on type or luminosity are manifestations of the same
phenomenon.   The {\it Third Reference Catalogue} description is in
the mid-to-upper range of the effect we see.  It is possible that high
luminosity 
early-type spirals have less obscuration than late types of the same
luminosity, as the {\it Third Reference Catalogue} corrections would
suggest.  The statistics with our sample are too slim to test this
possibility with only $\sim 25\%$ of our spirals typed S$a-b$.
From our limited information, we concur with the {\it Third Reference
Catalogue} that extinction appears to be negligible in S0 systems.

Bottinelli et al. (1995)\markcite{b95} find a slightly larger
$\gamma_B=1.67$ for pure disk systems, galaxies that would be typed
S$c-d$.  These corrections are substantial and correspond to
relatively opaque systems.   Bottinelli et al. (1995)\markcite{b95}
argue that $\tau_B\sim1$ as far out in the galaxies as radii
corresponding to a $B$ surface brightness of 25 mag arcsec$^{-2}$.  As
a consequence, they argue, diameters at that isophote are insensitive
to inclination.  Our results seem to be in reasonable agreement with
both Bottinelli et al. and the {\it Third Reference Catalogue}
although the details are difficult to compare because we formulate the
dependencies differently.  We would need a bigger sample to look for 
type variations. 

Considerable recent work has been done at $I$ band.
Han (1992)\markcite{h92} used the parameterization
$A_I^{i-0}=\gamma_I{\rm log}(a/b)$ with $\gamma_I=0.90$ for types
S$bc$-S$c$ and the lesser  values of $\gamma_I=0.73$ for types
S$0/a$-S$b$ and $\gamma_I=0.51$ for types S$cd$ and later.  These
results are in the upper half of our range and are a plausible
translation of our luminosity effect to a type effect.
Giovanelli et al. (1994)\markcite{g94}, in a first paper on
extinction, used the same analytic form and found 
$\gamma_I=1.05$, near the top end of our range.  In passing, we mention
a correction with a different formulation,
$A_I^{i-0}=1.37(1-b/a)$, given by Bernstein et al.
(1994)\markcite{bg94}.  This formulation gives much
larger adjustments over the full range of inclinations from edge-on to
face-on, and is somewhat larger even from $45\deg$ to $90\deg$, the 
range over which it
was defined.  The Bernstein et al. algorithm gives a poor fit toward
face-on, but otherwise all the 
descriptions compared here give consistent results within reasonable
uncertainties. 

In a follow-up paper,  Giovanelli et al. (1995)\markcite{g95} gave
attention to luminosity dependencies and it is this work that provides
the best possibility of a close comparison with the current results.
Data plotted in Fig.~7 of  Giovanelli et al. (1995)\markcite{g95} are
replotted as circles in the $I$ panel of Fig.~3 here, with an $0.48^m$
increase in the absolute magnitude scale to shift the distance scale
from H$_0=100$ \kms\ Mpc$^{-1}$ to H$_0=80$, so as to be consistent
with our Pisces distance.  The circles are filled or open as in the
Giovanelli et al. figure.  The error bars have not been carried over
but they are comparable to those associated with the filled squares.

The trends in the two data sets are similar, though Giovanelli et
al. find somewhat larger extinction at a given luminosity than is
found in the present analysis.  The offset is too large to be
attributed to magnitude differences due to a zero-point difference
in distances.  Perhaps the greater extinction found by Giovanelli et al.
is a manifestation of a type dependence separate from the luminosity 
dependence.  The Giovanelli et al. sample is dominantly composed of
late spirals which are the types suspected to have the highest extinction.
Our `complete' spiral sample is diluted by earlier types which plausibly
have less extinction at a given luminosity.  

The solid straight line in Fig.~3
is a least-squares fit to our data with weights dependent on the
inverse-square of the errors.
The dashed line is a similar fit to
the Giovanelli et al. material.  The two fits are formally
inconsistent.  The dotted line bisects the two fits and is equivalent
to a fit that gives equal weight to each of the two data sets.  This
line is marginally consistent as a fit to each data set.

We have no basis for a preference of one result over the other.  The
data providing the open circles from Giovanelli et al. are less
trustworthy because they are derived from a test that required
corrections to dimensions as well as magnitudes.  The filled circles
result from the same test that gives our filled triangles.  Our test
based on the color--magnitude correlation, that gives our filled
boxes, is intrinsically a more sensitive test.  However Giovanelli et
al. draw on a much larger sample of 2816 galaxies.  Hence they have
been able to use 10 bins instead of our 3 or 4.  We have a better
test, use galaxies of all inclination, and have a complete sample to
absolute magnitude limits but Giovanelli et al. have a lot more
galaxies.  Hence, we propose to use the description that bisects our
two results.  With more data, it might ultimately prove possible to 
provide a more refined description of the luminosity dependence of 
extinction, with type, perhaps, as a second parameter.

This conclusion with the $I$ band results raises the quandry that if
we advocate greater extinction than our $I$ material suggests then
what should we suppose is going on at $B$ and $R$?  Tentatively, we
suppose that if Giovanelli et al. had $B$ and $R$ data for their large
sample then they would have found greater extinction than we find.
The dotted lines in the $B$ and $R$ panels assume augmentations
above the solid lines such that at the luminosity of the brightest galaxy
in our sample $\gamma_{\lambda}$ is enhanced by 2\% over what we find 
and such that 
$\gamma_{\lambda}$ goes to zero $1^m$ 
fainter than found with our data.  These conditions are roughly
consistent with what is found at $\lambda=I$.  

Our comparisons would not be complete without mention that Willick
et al. (1996)\markcite{w96} explicitely discount a luminosity
dependence for extinction corrections.  They base their claim on
an analysis of the Mathewson, Ford, \& Buchhorn (1992)\markcite{mfb}
data set at $I$ and also their own material at $r_{gunn}$ (for which
Courteau (1996)\markcite{c96} finds $\gamma_r=0.95$).  Their analysis
is more indirect, asking if there are residuals in distance estimates
as a combined function of inclination and luminosity.  They find no
such correlation in residuals but the plots have a lot of scatter.
Our tests and those by Giovanelli et al. (1995)\markcite{g95} are
more direct and less buried in other components of scatter.

The extinction in the faintest galaxies in our sample is so low that
it is not significantly above our
measurement capability though other authors find evidence
for absorption at low luminosities or late types.  Our complete sample
in UMa extends roughly as faint as the Small Magellanic Cloud.  The
other data sets that have been discussed draw from galaxies at larger
redshift and contain relatively few galaxies as faint.  The status of
extinction in low luminosity galaxies is not yet on a good footing.

The equations that describe the short-dashed lines in Fig.~3 and, hence, 
represent the corrections we advocate are (with $h_{80}={\rm H}_0/80$):
\begin{equation}
\gamma_B = -0.35 (15.6 + M_B^{b,k,i} + 5{\rm log}h_{80})
\end{equation}
\centerline
{or $\gamma_B=0$ if $M_B^{b,k,i}>-15.6^m$,}
\begin{equation}
\gamma_R = -0.24 (16.2 + M_R^{b,k,i} + 5{\rm log}h_{80})
\end{equation}
\centerline
{or $\gamma_R=0$ if $M_R^{b,k,i}>-16.2^m$,}
\begin{equation}
\gamma_I = -0.20 (16.9 + M_I^{b,k,i} + 5{\rm log}h_{80})
\end{equation}
\centerline
{or $\gamma_I=0$ if $M_I^{b,k,i}>-16.9^m$,}
\begin{equation}
\gamma_{K^{\prime}} = -0.045 (18.3 + M_{K^{\prime}}^{b,k,i} + 5{\rm log}h_{80})
\end{equation}
\centerline
{or $\gamma_{K^{\prime}}=0$ if $M_{K^{\prime}}^{b,k,i}>-18.3^m$.}

The adjustments specified by the absorption model that have been discussed
have been applied
to the color--magnitude diagrams that are shown in Figure~4.  The
luminosity 
amplitude dependency illustrated by the short-dashed lines in Fig.~3 is applied.
Hence the faintest galaxies have little correction.  Because of this
luminosity dependence to the corrections, the color--magnitude
correlations become very steep.  As a consequence, these relationships
are not useful as distance estimators, contrary to the hopes of
Tully, Mould, \& Aaronson (1982)\markcite{tma}.  However the weak
dependence of the color-magnitude relations on distances means that 
relative uncertainties in the
distances of UMa and Pisces are not important for the present analysis.  
The rms scatter about
the regression (errors in color) are $\pm 0.40^m$ in $B-$\k', $\pm
0.25^m$ in $R-$\k', and $\pm 0.23^m$ in $I-$\k'.  The scatter is
greatest at the faint luminosities.

The 10 galaxies typed earlier than S$a$ are located in Fig.~4 with
special star symbols.  These galaxies have not received corrections for
obscuration.  They are redder than the mean of the spirals, lying at
the boundary of the scatter displayed by the spirals.

\section{Comparison between the Different Wavelength Bands}

Eq.~(3)-(5) imply $A_R/A_B=0.72$ and 
$A_I/A_B=0.59$ which are larger values than anticipated
if the Galactic reddening law along a
fixed path were applicable.  The relative Galactic fixed path reddening
values are 0.62 and 0.41, respectively (Cardelli, Clayton, \& Mathis 1989
\markcite{ccm}; but for $R,I$ filters in the Cousins, not Johnson, system
-- our filter central wavelengths are 6470\AA\ and 8320\AA, respectively).
Han (1992)\markcite{h92a} has correctly 
pointed out that the reddening should not fall as quickly toward the
infrared as would be anticipated from the Galactic reddening law
because path lengths are not the same at all wavelengths in the
present situation.  The Galactic law pertains to the obscuration
between the observer and a fixed source.  In the present situation,
the observer is looking to different distances at different
wavelengths along those lines-of-sight that become optically thick.
The observer will look deeper into nebulosities at $I$ than at $B$ but many
sight-lines will still be terminated by large optical depths.  

By extension it can be expected that obscuration at \k'\ is not entirely
negligible although it is too little to be measured directly by the
tests at our disposition.  From the Galactic law one anticipates
$A_{K^{\prime}}/A_B=0.085$.  However the Han effect must be at play and from
the enhancement of $A_I/A_B$ above the Galactic law value it can be 
concluded that $A_{K^{\prime}}/A_B>0.12$.  For our purposes, we will tentatively
accept $A_{K^{\prime}}/A_B\sim0.15$.  Corrections at \k'\ can reach $\sim0.3^m$
for the most luminous, highly inclined galaxies.

\section{Transposition of Linewidths for Luminosities}

The formulation of extinction corrections in terms of luminosities is
of limited utility for our primary interest, which is the use of
luminosity--linewidth correlations as distance estimators.  Absolute
luminosities are to be an output of the process so they cannot also be
an input.  However, the correction can work if it is based on the
value of the distance independent variable.  For our interests, the
distance independent variable is the HI profile linewidths
or some other measure of the rotation rate.  A similar motivation pushed
Giovanelli et al. (1997)\markcite{g97} to express absorption corrections 
in terms of their version of the rotation parameter.

Extinction corrected luminosity--linewidth correlations are
seen in Figure~5.  The straight lines illustrate linear regressions
with uncertainties in linewidths.  These fits are described by the
following equations, where the linewidth parameter, $W_R^i\simeq 2V_{max}$,
was discussed by Tully \& Fouqu\'e (1985)\markcite{tf85}:
\begin{equation}
M_B^{b,k,i} = -20.04 - 7.79 ({\rm log} W_R^i - 2.5)
\end{equation}
\begin{equation}
M_R^{b,k,i} = -21.09 - 7.96 ({\rm log} W_R^i - 2.5)
\end{equation}
\begin{equation}
M_I^{b,k,i} = -21.54 - 8.17 ({\rm log} W_R^i - 2.5)
\end{equation}
\begin{equation}
M_{K^{\prime}}^{b,k,i} = -23.16 - 8.73 ({\rm log} W_R^i - 2.5)
\end{equation}

As an aside, it can be noted that the luminosity--dependent extinction
corrections cause the bluer bands to be steeper than they appear in
the absence of such a dependency in the corrections.  Hence, there is
a much weaker color term in the slope than has been found in the past.
Also, the differential correction with luminosity acts to eliminate
curvature in the logarithmic luminosity--linewidth correlations.

The relations between $\gamma_{\lambda}$ and $M_{\lambda}$ presented 
in Eq.~(3)-(6) can now be recast to replace $M_{\lambda}$ terms
with $W_R^i$ terms.  The extinction parameter $\gamma_{\lambda}$ can go to zero but 
cannot go negative.
\begin{equation}
\gamma_B = 1.57 + 2.75 ({\rm log} W_R^i - 2.5)
\end{equation}
\begin{equation}
\gamma_R = 1.15 + 1.88 ({\rm log} W_R^i - 2.5)
\end{equation}
\begin{equation}
\gamma_R = 0.92 + 1.63 ({\rm log} W_R^i - 2.5)
\end{equation}
\begin{equation}
\gamma_{K^{\prime}} = 0.22 + 0.40 ({\rm log} W_R^i - 2.5)
\end{equation}
Fortunately, absorption values 
$A_{\lambda}^{i-0}=\gamma_{\lambda}{\rm log}(a/b)$ are stabilized
against inclination uncertainties because an error drives the 
$\gamma_{\lambda}$ and ${\rm log}(a/b)$ terms in offsetting senses.

The properties of the luminosity--linewidth correlations and the
measurement of distances are not our direct concern in this paper, but the
fits in Fig.~5 do strongly constrain the relative distances of the
Ursa Major and Pisces clusters.  The best distance is given by
minimizing the residuals of each cluster separately to the mean line
and taking the average of the four passbands.  $B$ is given a relative
weight of 0.7, in proportion to the square of the increased dispersion
in that band.  The relative distance modulus difference (Pisces--UMa)
is $2.55^m\pm0.10^m$.  The relative moduli in the different
bands agree with that value with $\pm 0.03^m$ rms dispersion.
The rms scatter about the
mean relationships is $0.42^m-0.43^m$ at $R$, $I$, and \k', and $0.55^m$
at $B$.  The uncertainty of $0.10^m$ in the distance estimate is the
quadrature sum of the separate standard deviations of the two clusters
in {\it each} of the $R$, $I$, and \k' bands.

If the distance modulus of the Ursa Major Cluster is taken to be 31.33
(Tully 1997) then Pisces is at 33.88, or 60 Mpc.  The cluster is
given a velocity in the frame of the cosmic microwave background of
$V_{CMB}=4771$ \kms\ by Han \& Mould (1992)\markcite{hm}.  The
inferred value for the Hubble Constant is $H_0=80$ \kms Mpc$^{-1}$.

\section{Conclusions}

Though our sample is moderate in number (pared to 87 spirals in the
final analysis) it is reasonably well defined in terms of completion
limits and we have photometric images in four bands ranging from $B$,
where there is considerable absorption, to \k', where there is
minuscule absorption.  Inclination 
effects of the sort attributable to extinction are seen in the scatter
of linewidth--magnitude and color--magnitude correlations.  More
edge-on galaxies are fainter and redder in the mean.  Plots of
deviations from the mean as a function of inclination provide a
quantitative measure of the amplitude of absorption.  Similar results
are found whether the residuals in linewidth--magnitude or
color--magnitude relations are considered.  The statistical
significance of results are better with the color--magnitude analysis
so we concentrate on those tests.

The claim by Giovanelli et al. (1995)\markcite{g95} is confirmed that
the extinction in spirals is luminosity dependent.  Galaxies with
$M_{K^{\prime}}^{b,k,i}$ reaching $-25^m$ ($M_B^{b,k,i}\sim-22^m$) have up to
$\sim 1.7^m$ of differential absorption between face-on and edge-on
($1.3^m$ at $R$, $1.0^m$ at $I$, $0.25^m$ inferred at \k').  However
the amplitude of extinction drops off rapidly with decreasing
luminosity, becoming unmeasurable at $M_{K^{\prime}}^{b,k,i}>-20^m$
($M_B^{b,k,i}\ga-17^m$).  The corrections reach higher values than 
we entertained in the past (Tully \& Fouqu\'e 1985)\markcite{tf85}.  
We arrived at 
our old corrections from data on relatively nearby galaxies which have 
low luminosities in the mean.  Samples such as those studied by
Giovanelli et al. (1994,1995) have larger mean redshifts, hence 
larger mean luminosities, and it can now be understood why such
samples lead to larger absorption corrections.
Extinction corrections have been formulated in terms of the
distance independent variable, linewidths, so they can be
applied to distance estimators.  The data assembled in this paper 
suggests $H_0=80$ \kms Mpc$^{-1}$.

\bigskip
Sadly, Peter Witchalls passed away a few days after this paper was
submitted for publication.

\clearpage

\begin{figure}
\caption{
Color--magnitude diagrams for the combined Ursa Major and Pisces
samples before corrections for extinction as a function of
inclination.   The Ursa Major galaxies are identified by circles and
the Pisces galaxies are identified by squares.  Filled symbols
correspond to systems more edge-on than $60\deg$ while open symbols
correspond to systems more face-on than this limit.  The conversion
from apparent to absolute magnitudes assumes the Ursa Major galaxies
are at a modulus of $31.33^m$ and the Pisces galaxies are at a modulus
of $33.88^m$.  The lines are regressions with uncertainties in colors.
The three panels illustrate the $B-$\k', $R-$\k', and $I-$\k'
dependencies, respectively.
} \label{1}
\end{figure}

\begin{figure}
\caption{
Deviations in color from the mean relations of Fig.~1 as a function of
axial ratio $b/a$.  The top row of panels illustrate deviations at
$B-$\k', the middle row illustrates deviations at $R-$\k', and the
bottom row illustrates deviations at $I-$\k'.  The left column of
panels isolates the fraction of the sample with $M_{K^{\prime}}<-23.5^m$
(9 galaxies in Ursa Major and 12 galaxies in Pisces), the panels second 
from left show data for galaxies with $-23.5^m<M_{K^{\prime}}<-22.5^m$
(11 galaxies in Ursa Major and 11 galaxies in Pisces), the panels second 
from right show data for galaxies with $-22.5^m<M_{K^{\prime}}<-21^m$
(13 galaxies in Ursa Major and 11 galaxies in Pisces), and the right
set of panels shows data for galaxies with $-21^m<M_{K^{\prime}}<-19.2^m$
(20 galaxies in Ursa Major and none in Pisces).  {\it Filled circles:} 
Ursa Major; {\it open circles:} Pisces.  The solid curve in
each panel gives the best fit for the analytic expression
$\gamma_\lambda {\rm log}(a/b)$ with the requirement that the dependence on
$b/a$ be nullified.  
}\label{2}
\end{figure}

\begin{figure}
\caption{
Dependence of the extinction amplitude parameter $\gamma_\lambda$ on
absolute magnitude.  Data are presented for $B,R,I$ bands in the three
separate panels.  The filled squares correspond to values of
$\gamma_\lambda$ consistent with the solid curve fits in Fig.~2 to
deviations from the mean color relations as a function of $b/a$.  The
filled triangles correspond to equivalent information derived from
deviations from luminosity--line profile width correlations as a
function of $b/a$.  The small circles in the $I$ panel are data taken
from Giovanelli et al. (1995), with open and closed
symbols identifying results from the two separate methods discussed in
that reference.  The dashed straight line in the $I$ panel is a least
squares fit to the Giovanelli et al. data (errors in $\gamma$).  The
solid straight lines in this and the other panels are least squares
fits to the data points of the present paper.  The dotted straight
line in the $I$ panel gives equal weight to the old and new data.  The
dotted lines in the $B$ and $R$ panels are offset from the solid lines
that correspond to the offset of the dotted line from the solid line
in the $I$ panel.
}\label{3}
\end{figure}

\begin{figure}
\caption{
Color--magnitude diagrams corrected for absorption effects as a function
of inclination.  {\it Filled circles:} Ursa Major; {\it open circles:}
Pisces; {\it stars:} types earlier than S$a$.  Straight lines are 
regressions with errors in colors to types S$a$ and later.
}\label{4}
\end{figure}

\begin{figure}
\caption{
Luminosity--linewidth relations in the four passbands after corrections for
inclination.  {\it Filled circles:} Ursa Major; {\it open circles:}
Pisces.  Straight lines are regressions with errors in linewidths.
}\label{5}
\end{figure}

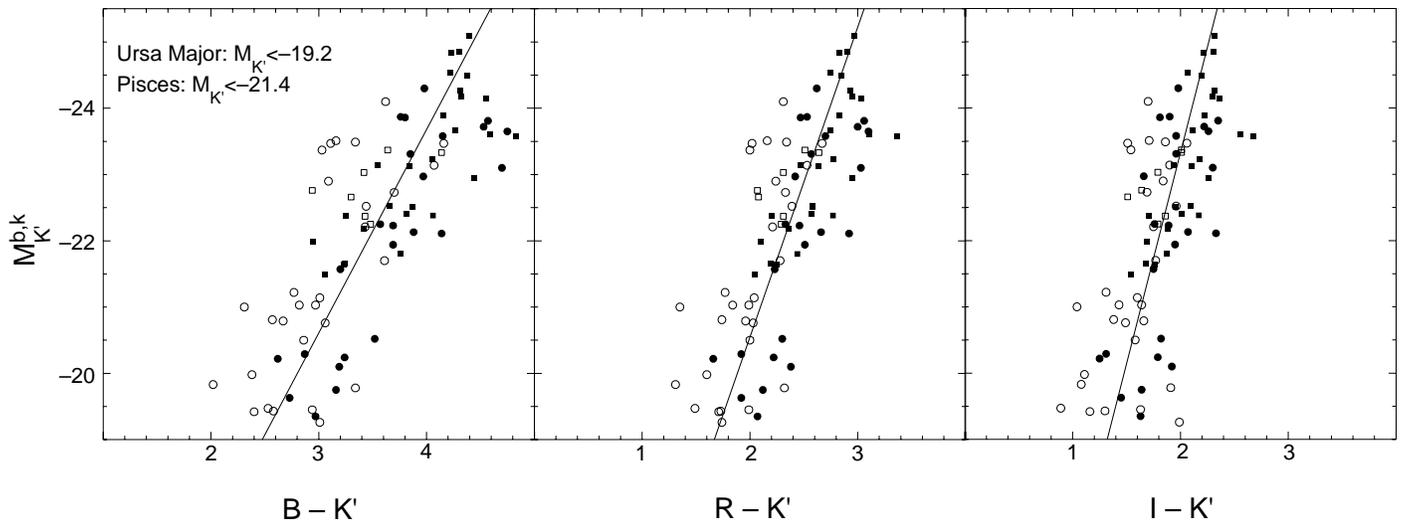
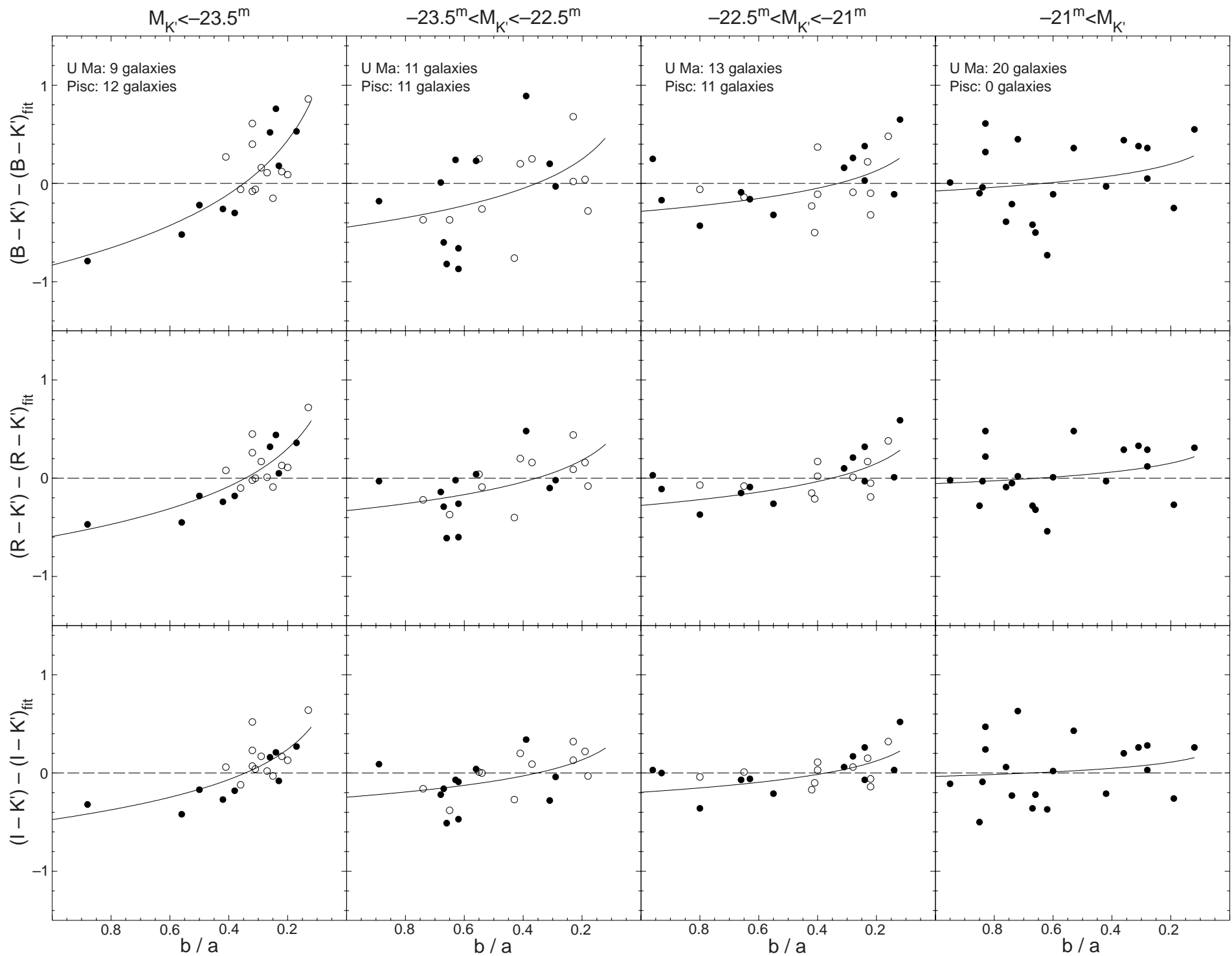
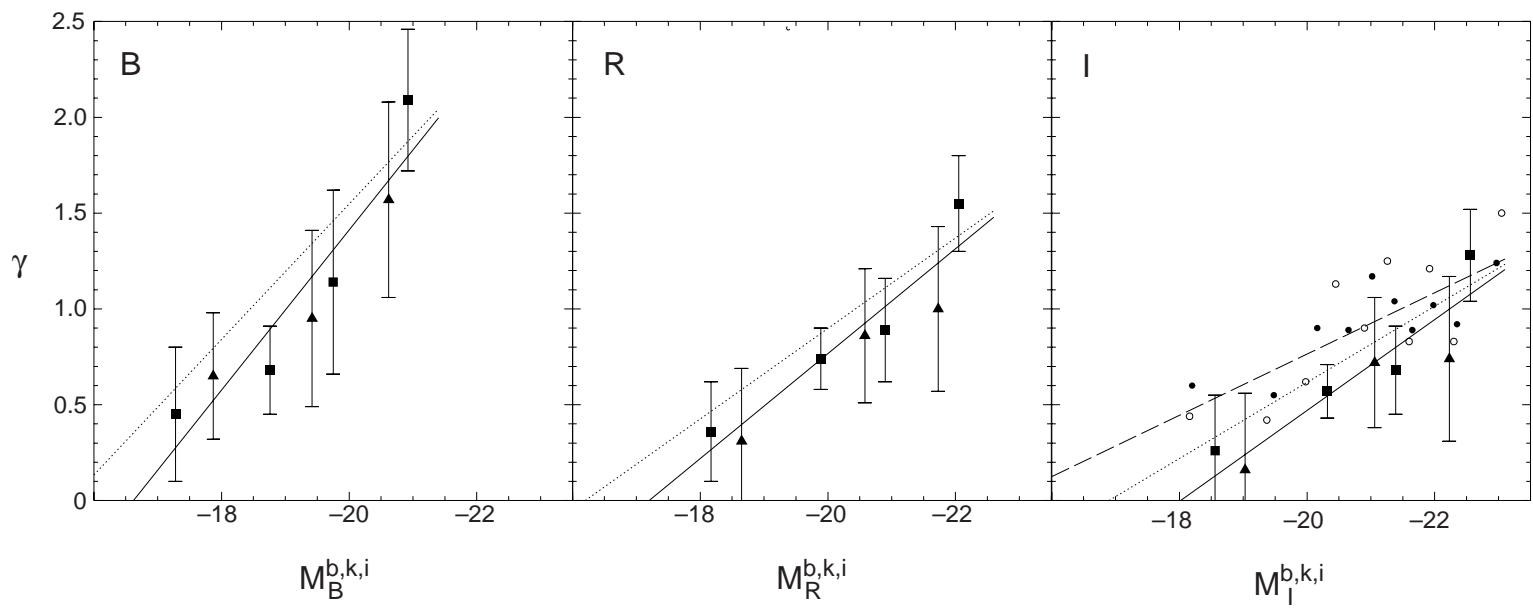
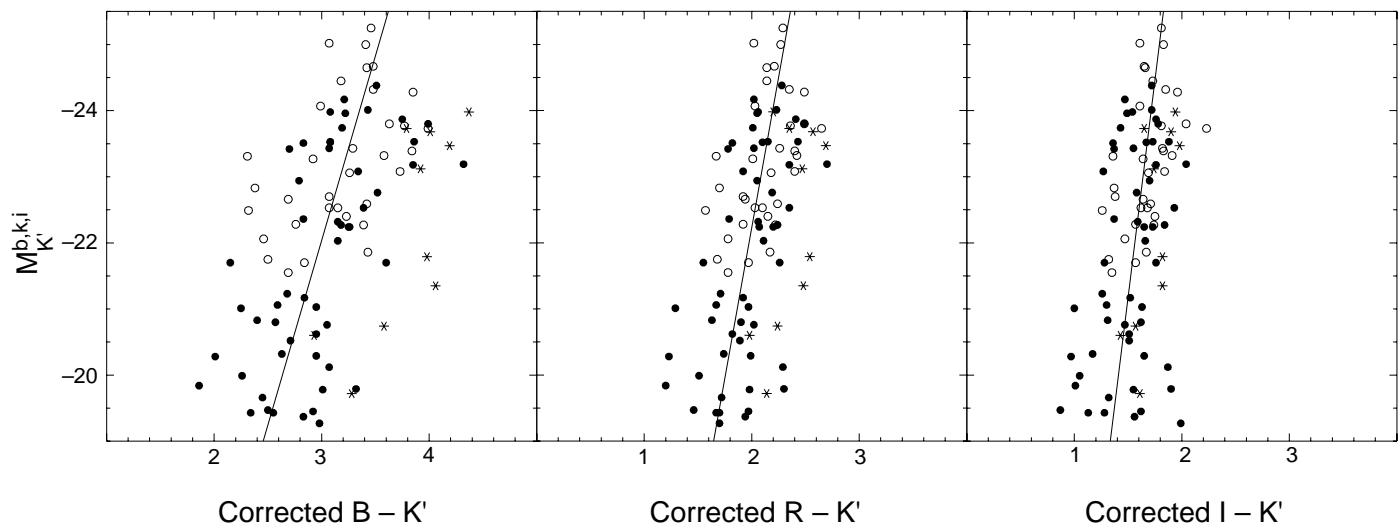
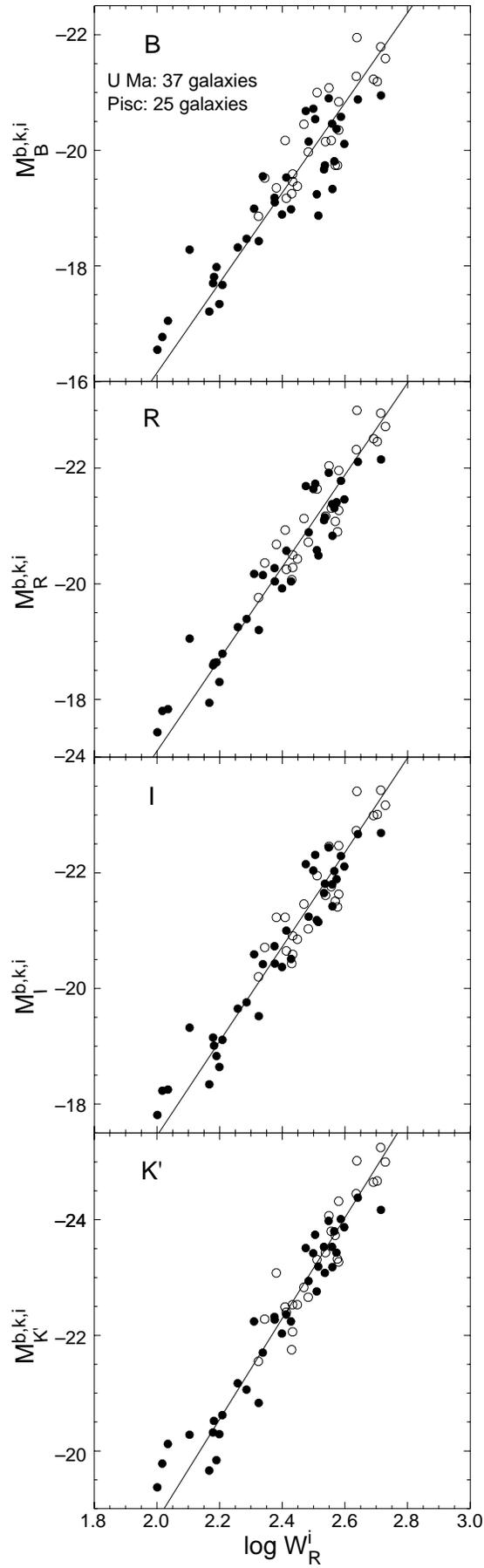
\end{document}